\documentclass[preprint,11pt,number,3p,times]{elsarticle}
\usepackage[latin1]{inputenc}

\bibstyle{elsarticle-num}

\usepackage[centertags]{amsmath}
\usepackage{amssymb}
\usepackage{hyperref}
\hypersetup{
    bookmarks=false,      % show bookmarks bar?
    pdfstartview={FitH},  % fits the width of the page to the window
    colorlinks=true,      % false: boxed links; true: colored links
    linkcolor=black,      % color of internal links
    citecolor=blue,       % color of links to bibliography
    filecolor=blue,       % color of file links
    urlcolor=blue         % color of external links
}
\usepackage{color}
\usepackage{epsfig}
\usepackage{hyperref}
\usepackage{varioref}
\usepackage{scrtime}
\usepackage{bbm}
\usepackage{longtable}
\usepackage{lscape}

\usepackage[usenames,dvipsnames,table]{xcolor}
\usepackage{fancyvrb}

\newcommand{\code}[1]{\small{\tt #1}\normalsize}
\newcommand{\orbifolder}{\code{orbifolder}}
\newcommand{\myvspace}{\vspace{0cm}}

\newcommand{\Id}{\ensuremath{1\!\!1}}
\newcommand{\SO}[1]{\ensuremath{\mathrm{SO}(#1)}}
\newcommand{\SU}[1]{\ensuremath{\mathrm{SU}(#1)}}
\newcommand{\U}[1]{\ensuremath{\mathrm{U}(#1)}}
\newcommand{\E}[1]{\ensuremath{\mathrm{E}_{#1}}}
\newcommand{\Z}[1]{\ensuremath{\mathbbm{Z}_{#1}}} % z_N ->\Z{N}

\newcommand{\bsb}[1]{\ensuremath{\boldsymbol{\overline{#1}}}}

\newcommand{\maR}{\ensuremath{\mathbbm{R}} }
\newcommand{\maZ}{\ensuremath{\mathbbm{Z}} }
\newcommand{\maQ}{\ensuremath{\mathbbm{Q}} }
\newcommand{\x}{\ensuremath{\times}}

\journal{Computer Physics Communications}

\begin{document}

\begin{frontmatter}

%\titlehead{
\begin{flushright}
\normalsize{LMU-ASC 51/11}\\
\normalsize{TUM-HEP 818/11}\\
\normalsize{DESY-11-187}\\
\normalsize{LPSC-11220}
\end{flushright}
%}

\title{\textbf{The Orbifolder:}\\
A Tool to study the Low Energy Effective Theory of Heterotic Orbifolds}

\author[bonn]{H.P.~Nilles}
\author[mexico]{S.~Ramos-S\'anchez}
\author[desy,tum,lmu]{P.K.S.~Vaudrevange}
\author[lpsc]{A.~Wingerter}

\address[bonn]{Bethe Center for Theoretical Physics and Physikalisches Institut der Universit\"at Bonn,
  Nussallee 12, 53115 Bonn, Germany}
\address[mexico]{Department of Theoretical Physics, Physics Institute, UNAM, Mexico D.F. 04510, Mexico}
\address[desy]{Deutsches Elektronen-Synchrotron DESY, Notkestra\ss e 85, 22607 Hamburg, Germany}
\address[tum]{Physik-Department T30, Technische Universit\"at M\"unchen, James-Franck-Stra\ss e, 85748 Garching, Germany}
\address[lmu]{Arnold-Sommerfeld-Center for Theoretical Physics, Theresienstra\ss{}e 37, 80333 M\"unchen, Germany}
\address[lpsc]{Laboratoire de Physique Subatomique et de Cosmologie, UJF Grenoble 1, CNRS/IN2P3, INPG,\newline 53 Avenue des Martyrs, F-38026 Grenoble, France}

\begin{abstract}
The \orbifolder{} is a program developed in \code{C++} that computes and 
analyzes the low-energy effective theory of heterotic orbifold compactifications. 
The program includes routines to compute the massless spectrum, to identify 
the allowed couplings in the superpotential, to automatically generate large 
sets of orbifold models, to identify phenomenologically interesting models (e.g. 
MSSM-like models) and to analyze their vacuum-configurations.
\end{abstract}

\begin{keyword}
Orbifold, string compactification, extra dimensions, particle spectrum, MSSM
\end{keyword}
\end{frontmatter}

\section*{Program Summary}
\noindent{\em Program title:} \orbifolder{}\\
{\em Program obtainable
  from:} {\tt http://projects.hepforge.org/orbifolder/}\\
{\em Distribution format:}\/ tar.gz\\
{\em Programming language:} {\code{C++}}\\
{\em Computer:}\/ Personal computer\\
{\em Operating system:}\/ Tested on Linux (Fedora 15, Ubuntu 11, SuSE 11)\\
{\em Word size:}\/ 32 bits or 64 bits\\
{\em External routines:}\/ None\\
{\em Dependencies:} Boost, GSL \\
{\em Typical running time:}\/ Less than a second per model.\\
{\em Nature of problem:}\/ Calculating the low energy spectrum of heterotic orbifold compactifications.\\
{\em Solution method:}\/ Quadratic equations on a lattice; representation theory; polynomial algebra. \\

%%%%%%%%%%%%%%%%%%% NEW SECTION %%%%%%%%%%%%%%%%%%%
\clearpage
\newpage
\section{Introduction}

String theory is a candidate for a consistent unified quantum theory
of gravity and gauge interactions and could thus provide us with an ultraviolet 
completion for models of particle physics. The search for 4-dimensional
string vacua resembling the standard model (SM) (or its supersymmetric 
extension (MSSM)) is therefore one of the central questions in string 
theory research. Over the years, a wide landscape of 4-dimensional string
models has emerged and it remains to be seen how particle physics 
phenomena can be incorporated within the scheme of string theory. 
Some hints point to a unification of gauge couplings within the framework of exceptional
groups (as e.g. $\E{8}$) but a direct road from  strings to particle 
physics has not yet been identified. It is perhaps the time to step 
back, collect and classify existing model constructions and try to 
identify properties relevant for a description of nature.

Here we present and analyze a specific approach that was studied already in the 
1980s and has led to interesting results since then: orbifold 
compactification~\cite{Dixon:1985jw,Dixon:1986jc,Ibanez:1986tp} of the 
heterotic strings~\cite{Gross:1984Dd,Gross:1985fr}. The reason for the success 
of this approach is ``computability'' paired with geometric intuition. Exact 
tools of conformal field theory are here at our 
disposal~\cite{Dixon:1986qv,Hamidi:1986vh} that are usually not available in 
approaches based on compactification on smooth manifolds. Besides, in 
particular $\E{8}\times \E{8}$ as a gauge group allows a perturbative inclusion of 
the standard model gauge group (as well as possibly grand unification).

A toroidal orbifold is flat with the exception of a number of 
fixed points or fixed tori. It gives rise to a picture called 
the heterotic brane world scenario~\cite{Forste:2004ie,Kobayashi:2004ya,Buchmuller:2005jr}: fields can either live in 
the 10-dimensional bulk (untwisted sector) or can be localized 
at these fixed points or fixed tori (twisted sectors). The relative 
location of these fields as well as the local gauge structure 
determines many properties of the 4-dimensional string vacua and 
is the source of geometric intuition for model building. The 
orbifold point is a point of enhanced symmetries (in the moduli 
space of compactification) and those symmetries might be relevant 
for a description of nature. Models of particle physics seem to 
require many (discrete) symmetries, e.g. flavour symmetries or 
symmetries to prevent too fast proton decay. Some of these 
symmetries could be slightly broken to provide us  with small 
parameters relevant for the description of hierarchies as 
e.g. observed in the spectrum of quark and lepton masses. This 
supports our hope that orbifold compactification is not just an 
approximation with increased ``computability'', but that it provides 
a realistic compactification: nature might have chosen to live
close to the orbifold point with enhanced symmetries.

Explicit orbifold model constructions in the last 5 to 10 years have been 
extremely successful~\cite{Choi:2006qh,Nilles:2008gq,Vaudrevange:2008sm,RamosSanchez:2008tn} (see~\cite{Quevedo:1996sv,Bailin:1999nk} for earlier reviews). 
Many of the properties of particle physics 
can be incorporated in the scheme. This includes grand unification, 
gauge-Yukawa unification, satisfactory Yukawa textures, solutions 
to the $\mu$-problem, the creation of hierarchies and a successful 
incorporation of neutrino Majorana masses. Discrete symmetries are 
identified to solve the flavour  problem and to avoid too fast 
proton decay. These properties have been identified in the so called 
Minilandscape (based on the $Z_6$-II orbifold)~\cite{Lebedev:2006kn,Lebedev:2007hv,Lebedev:2008un}
and subsequent work~\cite{Lebedev:2006tr,Kim:2007mt,Buchmuller:2007zd,Dundee:2008ts,Kappl:2008ie,Nibbelink:2009sp,Hosteins:2009xk,Choi:2009jt,Blaszczyk:2009in,Lebedev:2009ag,Blaszczyk:2010db,Forste:2010pf,Parameswaran:2010ec,Kappl:2010yu}.

The search for such models requires a computer assisted scan that 
incorporates the rules for a consistent string theory construction 
(as e.g. modular invariance). The purpose of this work is to make 
the tools and techniques available to the public. We hope that this
will give more people the opportunity to contribute to this exciting 
field of model constructions.

We present the \orbifolder{}, a program developed in \code{C++} that 
allows the calculation of the low-energy spectrum of heterotic orbifold 
constructions. The program includes routines to compute the massless 
spectrum and to identify the allowed couplings of the superpotential. 
It allows the construction of arbitrary orbifold models, the identification 
of phenomenologically interesting models and a classification of their 
vacuum configurations.

The \orbifolder{} can be considered in some aspects as the stringy analog of 
programs such as \code{SoftSusy}, \code{SuSpect} and \code{SPheno}: The latter 
are devoted to the detailed computation of particle spectra, interactions and 
phenomenological quantities, using as input a high scale supersymmetric model 
and imposing low-energy constraints. Analogously, the \orbifolder{} takes as a 
starting point the 10D heterotic strings and, provided some geometric input 
describing the features of the six compactified dimensions, computes the 
(massless) spectrum, interactions and symmetries of the resulting lower-energy 
effective 4D field theory.

After a short introduction to heterotic orbifold compactification in Section 
\ref{sec:orbis}, we explain how to download and install the \code{C++} program 
in Section \ref{sec:download}. Section \ref{sec:howtorun} discusses the 
explicit recipe to run the program, while in Section \ref{sec:conclusions} we 
conclude and give an outlook for future research. Technical details are 
relegated to the appendices and to the webpage 
\cite[\S Complementary notes]{ORB11}.

%%%%%%%%%%%%%%%%%%% NEW SECTION %%%%%%%%%%%%%%%%%%%
\section{Heterotic Orbifold Compactifications}
\label{sec:orbis}

In this section we give a brief introduction to heterotic orbifolds, in order 
to introduce our notation and conventions used in the \orbifolder{}. For 
more details on orbifold compactifications, we refer to the 
reviews~\cite{Quevedo:1996sv,Bailin:1999nk,Choi:2006qh,Vaudrevange:2008sm,RamosSanchez:2008tn}.

In the context of heterotic string compactifications, we define an orbifold as 
the quotient of six-dimen\-sional real space $\mathbbm{R}^6$ divided by the 
so-called {\it space group} $S$, where the quotient is taken using the 
equivalence relation $X \sim gX$ for all $g\in S$ and $X \in \maR^6$. More 
specifically, the space group is chosen to consist of two parts: 
\begin{itemize}
\item discrete rotations that form the so-called {\it point group} $P$. For 
simplicity, we choose $P$ to be Abelian. To obtain $\mathcal{N} = 1$ 
supersymmetry in 4D, $P$ is either \Z{M} or \Z{M}\x\Z{N} generated by rotation 
matrices $\theta$ and $\omega$, where we use $\omega = 1$ for \Z{M}. These 
matrices can be represented by so-called {\it twist vectors} $v_1$ and $v_2$ 
that give the three rotational phases in the three complex planes and the sum 
over all entries integer to ensure $\mathcal{N} = 1$. For example, 
$v_1 = \left(0, \frac{1}{3}, \frac{1}{3},-\frac{2}{3}\right)$ and $v_2 = 0$ for 
the \Z{3} orbifold.
\item translations generated by the vectors $e_\alpha \in \maR^6$, for 
$\alpha=1,\ldots,6$. They form a 6D lattice denoted by $\Gamma$ and hence 
define a six-torus. Elements of $P$ must map the lattice $\Gamma$ to itself.
\end{itemize}
In detail, an element of the space group is of the form $g = \left(\theta^k 
\omega^l ,n_\alpha e_\alpha\right) \in S$, where $k,l\in\maZ$, $n_\alpha\in\maZ$ 
(or in some cases $n_\alpha\in\maQ$) and summation over $\alpha = 1,\ldots, 6$. It acts on 
$X\in\mathbbm{R}^6$ as $g X = \theta^k \omega^l X + n_\alpha e_\alpha$. Using 
these definitions, we can deal with all 6D Abelian and toroidal orbifolds 
including the cases of roto-translations and freely acting involutions (see 
e.g. Ref.~\cite{Donagi:2008xy}).

Due to modular invariance, the geometric action of the space group $S$ has to 
be embedded into the gauge degrees of freedom of the heterotic string, denoted 
by $X^I$, $I=1,\ldots,16$ in the bosonic formulation. We restrict ourselves to 
the case of shift embedding, where $\theta \hookrightarrow V_1$, $\omega 
\hookrightarrow V_2$ and $e_\alpha \hookrightarrow W_\alpha$ for $\alpha =
1,\ldots,6$. Then, the action of $S$ on $X \in \maR^6$ induces an action on 
$X^I$ as
\begin{equation}
\label{eq:gaugeembedding}
g\; X = \theta^k \omega^l X + n_\alpha e_\alpha \quad\Rightarrow\quad g\; 
  X^I = X^I + k V_1^I + l V_2^I + n_\alpha W_\alpha^I\;,
\end{equation}
where $I=1,\ldots,16$. The vectors $V_1$, $V_2$ and $W_\alpha$ are called 
{\it shifts} and {\it Wilson lines}, respectively. They are constrained by 
modular invariance \cite{Dixon:1986jc,Vafa:1986wx,Ploger:2007iq}, e.g.
\begin{equation}
M\,(V_1^2 - v_1^2) = 0\;\text{mod}\;2\quad\text{and}\quad 
N_\alpha W_\alpha^2 = 0\;\text{mod}\;2\;,
\end{equation}
where $M$ is the order of \Z{M} and $N_\alpha$ the order of $W_\alpha$. The 
combined group formed by the space group and its action on the gauge degrees of 
freedom is called the {\it orbifold group}.

From a model-building standpoint, we have now introduced all the input data to 
define a heterotic orbifold model: the space group $S$ (consisting of the point 
group $P$ and the lattice $\Gamma$) and its embedding as shifts $V_1$, $V_2$ 
and Wilson lines $W_\alpha$.

We close this section with a very brief summary of the construction of massless 
spectra of heterotic orbifolds. Given the input data, one distinguishes between 
two kinds of closed (and massless) strings on the orbifold: first of all 
ordinary closed strings, also called {\it untwisted strings}, being the 
remnants of the 10d \E{8}\x\E{8} or \SO{32} vector multiplet and the gravity 
multiplet. Secondly, there are closed strings from the {\it twisted sectors} 
which close only up to the action of the orbifold group. To construct them, one 
needs to identify the inequivalent non-trivial space group elements as the 
constructing elements of twisted strings: i.e. for $g\in S$ one can define a 
twisted boundary condition $X(\tau, \sigma + \pi) = g X(\tau, \sigma)$ on the 
string world-sheet. Then, using standard CFT techniques, the equations for 
massless right- and left-movers with boundary condition $g$ read
\begin{equation}
\label{eqn:massless}
\frac{\left(q + v_g \right)^2}{2} - \frac{1}{2} + \delta c = 0 \quad\text{and}
\quad \frac{\left(p + V_g \right)^2}{2} + \tilde{N} - 1 + \delta c = 0
\end{equation}
where $p$ is from the $\E{8}\times\E{8}$ or $\text{Spin}(32)/\Z{2}$ weight 
lattice, $q$ from the vector or spinor weight lattice of $\SO{8}$, $\delta c$ 
denotes the shift in the zero-point energy and $\tilde{N}$ the number operator 
of left-moving oscillators. Furthermore, we define the {\it local twist} $v_g = 
k v_1 + l v_2$ and the {\it local shift} $V_g = k V_1 + l V_2 + n_\alpha W_\alpha$. 
In the final step, one builds massless states as tensor products of massless 
right- and left-movers such that they are invariant under the full orbifold 
action, i.e.
\begin{equation}
\label{eqn:invariantstate}
|q + v_g\rangle_\text{R} \otimes |p + V_g\rangle_\text{L} \quad\text{or}\quad 
|q + v_g\rangle_\text{R} \otimes \tilde{\alpha} |p + V_g\rangle_\text{L}\;,
\end{equation}
where $\tilde{\alpha}$ denotes possible oscillator excitations. We define the 
shifted momenta $q_\text{sh} = q + v_g$ and $p_\text{sh} = p + V_g$, where 
$p_\text{sh}$ describes the transformation properties under gauge 
transformations. The states Eq.~\eqref{eqn:invariantstate} correspond to massless 
fields of the 4D effective field theory. They carry gauge charges (from 
$p_\text{sh}$), discrete $R$ charges, modular weights (from $q_\text{sh}$ and possible 
oscillator excitations) and discrete non-$R$ charges (from the constructing 
element $g \in S$).

%%%%%%%%%%%%%%%%%%% NEW SECTION %%%%%%%%%%%%%%%%%%%
\section{Download and Installation}
\label{sec:download}

The minimal requirements for compiling the \orbifolder{} are the Boost 
C++ Libraries version $\geq1.0$~\cite{boost:2011:reference} and the GNU 
Scientific Library (GSL) version $\geq1.9$ \cite{GSL}.
All components should come preinstalled on a standard Linux distribution. If 
this should not be the case, they can easily be installed. 

On a \code{yum}-based distribution like e.g.~Fedora, the command ``\code{yum 
-y install gsl gsl-devel boost boost-devel}'' will install the corresponding 
libraries (recommended). Alternatively, one can also directly install from 
source or use the other download options available 
\cite{boost:2011:reference,GSL}. 

The \orbifolder{} is free software under the copyleft of the \href{http://www.gnu.org/copyleft/gpl.html}{GNU General Public License} and can be downloaded from \cite{ORB11}:
\begin{center}
\href{http://projects.hepforge.org/orbifolder/}{http://projects.hepforge.org/orbifolder/}
\end{center}

To install the program, download the file \code{orbifolder-1.0.tgz} to a directory of your choice, open a shell and enter the following commands at the prompt:
\begin{Verbatim}[fontsize=\normalsize,numbers=left,xleftmargin=20pt,formatcom=\color{gray}]
tar xfvz orbifolder-1.0.tar.gz
cd orbifolder-1.0
./configure 
make
make install
\end{Verbatim} 
\label{verb:orbifoldinstallation}

Note that the version number ``1.0'' may change over time and should be substituted accordingly.
The first line unpacks the tar-ball and creates a subdirectory structure with the source 
code. The second line changes to the installation directory. The third line starts 
the configuration script that checks system requirements and generates the 
\code{Makefile}. The fourth line compiles the code, and finally the fifth line 
installs it on your system. After successful compilation and installation, the main program (named 
\code{orbifolder}) will be available in the current directory. We have disabled custom installation using the \code{--prefix} switch in the configuration script. The main program \code{orbifolder} can simply be copied to the directory of the user's choice by the standard shell commands. Detailed installation instructions can also be 
found in the README file in the installation directory and on the website 
\cite{ORB11}.

We have tested the installation process on
\begin{itemize}
\item a 32-bit system running Linux Ubuntu 11.04 with Boost 1.42 and GSL 1.14,
\item a 64-bit system running Linux SuSE 11 with Boost 1.36 and GSL 1.11,
\item a 64-bit system running Linux Fedora 15 with Boost 1.46 and GSL 1.14,
\end{itemize} and ascertained that our code compiles correctly. Should
there arise any problems during the installation, we request that the
user send us the file \code{config.log} and the output of the
\code{make} command by email
(\href{mailto:orbifolder@projects.hepforge.org}{orbifolder@projects.hepforge.org}).

%%%%%%%%%%%%%%%%%%% NEW SECTION %%%%%%%%%%%%%%%%%%%
\section{How to run the program}
\label{sec:howtorun}

There are three main ways to gain access to the \orbifolder{}: through the 
\code{prompt}, through a web interface and directly through the \code{C++} 
source code. In the following we will present them in detail.

%%%%%%%%%%%%%%%%%%% NEW SUBSECTION %%%%%%%%%%%%%%%%%%%

\subsection{The prompt}
\label{sec:prompt}

The \orbifolder{} can be controlled using a Linux-style command line called 
the \code{prompt}. The \code{prompt} offers an interactive access to almost all 
variables and functions of the \orbifolder{}. It has the structure of a 
file system where orbifold models appear as directories. In the following we 
explain how to start and use the \code{prompt}.

%%%%%%%%%%%%%%%%%%% NEW SUBSUBSECTION %%%%%%%%%%%%%%%%%%%

\subsubsection{How to get started}
\label{sec:howtogetstarted}

We begin with a small tutorial, see Tab. 4 in the additional material 
\cite[\S Complementary notes]{ORB11} for a sample input and output. In general, the 
\code{prompt} can be started using the command \code{./orbifolder} or 
\code{./orbifolder [model file]}. In the former case, no model is loaded 
automatically. In the latter case, the details of a model contained in the 
plain-text-based \code{model file} are loaded (Further properties of {\it model 
files} are explained in Section \ref{sec:modelfile}). As an example, run the 
program using the command
\begin{equation}
\text{\code{./orbifolder modelZ3.txt}}
\end{equation}
with parameter \code{modelZ3.txt} in order to load the standard embedding model 
of the \Z{3} orbifold from this file. Having started the program, one enters 
the \code{prompt} in its main directory \code{/>}. The command
\begin{equation}
\text{\code{dir}}
\end{equation}
lists all commands and subdirectories of the current directory. In our example, 
there is one subdirectory in the main directory \code{/>} called 
\code{Z3StandardEmbedding} which corresponds to the \Z{3} model loaded. In 
general, a (sub-) directory \code{A} can be accessed using the command \code{cd 
A}. In our example,
\begin{equation}
\text{\code{cd Z3StandardEmbedding}}
\end{equation}
results in the output \code{/Z3StandardEmbedding>} from where one can access, 
analyze or change the details of the \Z{3} standard embedding model. Again, 
type in the command \code{dir} to see all commands and subdirectories of the 
current directory. For all orbifold model directories there are five 
subdirectories, 
\begin{equation}
\text{\code{/model>},\; \code{/gauge group>},\; \code{/spectrum>},\; 
\code{/couplings>}\; and\; \code{/vev-config>},}
\end{equation}
containing commands of the respective category:
\begin{itemize}
\item \code{/model>}: Print and change the input data of the current orbifold 
      model. See~\ref{app:directorymodel}.
\item \code{/gauge group>}: Print details on the gauge group, change the 
      $\U{1}$ basis and find accidental $\U{1}$ symmetries. 
      See~\ref{app:directorygaugegroup}.
\item \code{/spectrum>}: Print details on the spectrum of massless fields. 
      See~\ref{app:directoryspectrum}.
\item \code{/couplings>}: Create and analyze the superpotential and the 
      resulting effective mass matrices. See \ref{app:directorycouplings}.
\item \code{/vev-config>}: Define and analyze various vev-configurations. 
      See~\ref{app:directoryconfig}. Each configuration is specified by the 
      distinction between observable and hidden sector of the gauge group and 
      the assignment of labels and vacuum expectation values to the fields 
      (labels are assigned in a subdirectory called \code{/labels>}, 
      see~\ref{app:directorylabels}).
\end{itemize}
\begin{table}[b]
\begin{center}
\begin{tabular}{|@{\;\small\tt}l|@{\normalsize}|l|}
\hline
command                   & description\\
\hline
\hline
dir                       & display commands and subdirectories of the current 
                            directory\\
cd A                      & change the current directory to \code{A} (if 
                            \code{A} exists)\\ 
cd ..                     & go back one directory\\
cd {\scriptsize$\sim$}    & go back to the main directory \code{/>}\\
exit                      & exit the program if no process is running; use the 
                            parameter\\
                          & \code{orbifolder} to enforce exit (also inside a 
                            script)\\
\hline
\end{tabular}
\end{center}
\caption{Some standard commands in the {\code{prompt}}.}
\label{tab:promptstandardcommands}
\end{table}
Again, one can access these directories using the \code{cd} command. For 
example, try \code{cd gauge group} to enter the subdirectory 
\code{/gauge group>} and use the commands \code{print gauge group} and 
\code{print simple roots} to see the gauge group ($\E{6} \times \SU{3} \times 
\E{8}$) and (a choice of) the corresponding simple roots. In order to go back 
one directory one uses the command \code{cd ..} at the prompt. Next, try 
the subdirectory \code{/spectrum>} and use the command \code{print summary} to 
get a summary table of all massless matter fields. The command 
\code{cd} {\scriptsize$\sim$} is used to go back to the main directory 
\code{/>}. Further standard commands of the \code{prompt} are described in 
Tab.~\ref{tab:promptstandardcommands}; see also~\ref{app:glossarycommands} for 
a glossary of commands. 

%%%%%%%%%%%%%%%%%%% NEW SUBSUBSECTION %%%%%%%%%%%%%%%%%%%

\subsubsection{How to create new orbifold models}
\label{sec:createneworbifolds}

Being in the main directory \code{/>} of the \code{prompt} new orbifold models 
can be accessed basically in three ways:
\begin{itemize}
\item \code{load orbifolds(Filename)}: Load orbifold models from a model file. 
      For example, load the \Z{6}--II orbifold MSSM of \cite{Buchmuller:2006ik} 
      using the command \code{load orbifolds(modelBHLR.txt)}.
\item \code{create orbifold(A) with point group(M,N)}: Create a \Z{M} or 
      \Z{M}\x\Z{N} orbifold named \code{A} by specifying $M$ and $N$ (set 
      $N=1$ for \Z{M}). After entering the new directory using \code{cd A}, 
      one is asked to specify more details on the model like shifts and Wilson 
      lines, see~\ref{app:directorymodel}.
\item \code{create random orbifold from(A)}. See below.
\end{itemize}
Orbifold models can be created randomly by using the command
\begin{equation}
\text{\code{create random orbifold from(A)}}
\end{equation}
in the main directory \code{/>}. The parameter \code{A} must be either the name 
of an existing (loaded or previously created) orbifold model or \code{*} for 
any existing orbifold model in the \orbifolder{}. The command starts a new 
process that runs in the background so that one can continue to work with the 
\code{prompt} (see~\ref{app:processes} for more details on processes). One can 
specify several optional parameters:
\begin{itemize}
\item \code{if(...)}: Specify the desired properties of the model: 
      \code{inequivalent} in order to choose only models with inequivalent 
      spectra and \code{SM}, \code{PS} or \code{SU5} for models with a (net) 
      number of \code{X} generations of Standard Model (\code{SM}), Pati-Salam 
      (\code{PS}) or $\SU{5}$ gauge group plus vector-like exotics, where 
      \code{X} is 3 by default and can be changed using the parameter 
      \code{Xgenerations}; c.f. the command \code{analyze config} 
      in~\ref{app:directoryconfig}.
\item \code{save to(Filename)}: Save the models with the desired properties to 
      a model file.
\item \code{use(1,1,0,1,...)}: Eight digits for two shifts plus six Wilson 
      lines; either 1 if the corresponding shift/Wilson line shall be taken 
      from model \code{A}, or 0 if it shall be created randomly.
\item \code{\#models(X)}: Define how many models (with the desired properties, 
      if specified) shall be created randomly. Use \code{\#models(all)} to 
      create as many models as possible. If \code{\#models(X)} is not used, 
      only one model shall be created.
\item \code{print info}: Print a short summary of the spectrum immediately when 
      a new model with the desired properties has been found.
\item \code{load when done}: Load the created models into the \orbifolder{} 
      after the process has finished.
\item \code{do not check anomalies}: Use this parameter to speed up the process.
\item \code{compare \#couplings of order(X)}: If only \code{inequivalent} 
      models are saved, this parameter refines the comparison between two 
      models: compare in addition the number of all superpotential couplings up 
      to the specified order \code{X}. Slows down the process considerably.
\end{itemize}

\myvspace
\paragraph{Examples} A typical example of how to use this command looks like
\begin{eqnarray}
&& \text{\code{create random orbifold from(Z3StandardEmbedding) if(inequivalent)}} \\
&& \qquad\text{\code{save to(Z3NewModels.txt) use(1,1,0,0,0,0,0,0) \#models(10) print info}} \nonumber
\end{eqnarray}
executed from the main directory \code{/>}. In this case a new process is 
started that constructs new \Z{3} orbifold models using both shifts (i.e. $V_2 
= 0$) but no Wilson lines from model \code{Z3StandardEmbedding}, saves only 
inequivalent models to a model file named \code{Z3NewModels.txt}, prints a 
brief summary of each new model and stops after creating ten inequivalent 
models. In a second example, ten random models of \code{SM} (Standard Model) 
type are created starting from the \Z{6}--II orbifold MSSM of 
\cite{Buchmuller:2006ik} by using the command
\begin{eqnarray}
&& \text{\code{create random orbifold from(Z6IIOrbifold\_BHLR) if(inequivalent SM)}} \\
&& \qquad\text{\code{save to(Z6IINewMSSMModels.txt) use(1,1,1,1,1,1,0,0) \#models(10)}} \nonumber \\ 
&& \qquad\text{\code{print info load when done}} \nonumber
\end{eqnarray}
in the main directory \code{/>}. The parameter \code{use(1,1,1,1,1,1,0,0)} 
specifies that only the Wilson lines $W_5$ and $W_6$ are created randomly, i.e. 
the shifts and other Wilson lines are taken from the original model. 
Furthermore, only \code{inequivalent} standard models (\code{SM}) are printed, 
saved to file \code{Z6IINewMSSMModels.txt} and finally loaded into the 
\orbifolder{} after the process has finished. Note that these MSSM models 
should be part of the Mini-Landscape \cite{Lebedev:2006kn,Lebedev:2008un}.

%%%%%%%%%%%%%%%%%%% NEW SUBSECTION %%%%%%%%%%%%%%%%%%%

\subsection{The web interface}

The \orbifolder{} can also be accessed through a user-friendly web interface.
The \code{orbifolder on-line} makes extensive use of the \code{prompt} 
explained in Section \ref{sec:prompt} rendering it available to any user with 
access to an internet browser, such as Mozilla Firefox version $\geq3.6$, 
Google Chrome version $\geq2.1$, and Internet Explorer $\geq8.0$. Consequently, 
the program is also available to users of all kinds of smartphones.

One of the advantages of the web interface is that one does not need to install 
any program on the local computer to be able to use most of the functions of 
the \orbifolder{}. Another advantage is that it can be executed from platforms 
that work with the most popular operating systems (without further auxiliary 
applications): Windows, Linux and Mac. 

On the less bright side, one shortcoming of this version is that it is not 
recommended to execute time-consuming instructions since short interruptions in 
the internet service may affect the results. Furthermore, a command running 
during more than 60 minutes is disabled automatically to avoid overload of the 
server. Finally, for security reasons, commands involving file manipulation are 
extremely limited. Specifically, the parameters and commands \code{ @begin/@end 
print to file(Filename)}, \code{save to(Filename)}, \code{load/save} \; 
\code{couplings(Filename)}, \code{load/save labels(Filename)} are disabled.

The \code{orbifolder on-line} can be used on our main page~\cite{ORB11}\\
\centerline{\href{http://projects.hepforge.org/orbifolder/}{\code{http://projects.hepforge.org/orbifolder/}}}
which redirects to any of our mirror servers.
To gain access, you must click on the link \code{orbifolder on-line}. This starts an \orbifolder{} session on the 
selected server. The new page consists of three parts:
\begin{itemize}
\item \code{History}: The result of the latest instructions is shown. The button \code{download history}
      provides an RTF file containing the full history, i.e. not only the result of the latest 
      instructions, but the result of all the commands used during the current session. The user can
      resize this window.
\item \code{List of Commands}: This is the input area, where the commands of the \code{prompt}
      are typed. To execute them, it is necessary to click on the button \code{execute commands}. An
      additional help in this section is the button \code{upload commands}. This button can be used when
      files (not larger than 100Kb) in plain-text format containing (lists of) admissible commands have 
      been prepared. The button \code{download commands} provides an RTF file containing the list of
      all the commands typed during the active \orbifolder{} session.

      Lists of useful commands and their use is provided in Section \ref{sec:prompt} and~\ref{app:glossarycommands}.
\item \code{Help}: The bottom part contains a list of all available commands 
      for the current directory. Each command in the displayed list is a link 
      to a more precise description of its use.
\end{itemize}

To terminate an \code{orbifolder on-line} session, it suffices to click on the 
upper button \code{EXIT}. The buttons provided on the resulting page allow the 
user to download the full history of the current session and the complete list 
of successfully executed commands.

Occasionally, errors in the input data may cause the \orbifolder{} to 
crash. In those cases, the web interface shall close your session, giving you 
the opportunity to download the list of instructions (button \code{download 
commands}) to be used if you restart the \orbifolder{}. Please, make sure 
that the downloaded list of commands does not contain the instructions that led 
to the failure of the program. We encourage users to contact the authors 
reporting any failure in the program, preferably by using the link 
\code{contact us} on the main page of the 
\href{http://stringpheno.fisica.unam.mx/orbifolder}{web interface}.

%%%%%%%%%%%%%%%%%%% NEW SUBSECTION %%%%%%%%%%%%%%%%%%%

\subsection[The \code{C++}\ source code]{The \texttt{\textbf{C++}} source code}

The \orbifolder{} is written in \code{C++}, distributed over several files. 
Many physical quantities, as briefly introduced in Section~\ref{sec:orbis}, 
have been encapsulated into classes, for example
\begin{itemize}
\item \code{CSpaceGroup} for the space group $S$ and \code{CSpaceGroupElement} 
      for its elements $g \in S$, 
\item \code{CTwistVector} for the twists $v_i$ and \code{CShiftVector} for the 
      corresponding shifts $V_i$, 
\item \code{CWilsonLines} for the set of six Wilson lines $W_\alpha$, 
\item \code{COrbifoldGroup} for the orbifold group, 
\item \code{CMasslessHalfState} for finding massless left- and right-movers, 
      i.e. solutions to Eq.~\eqref{eqn:massless},
\item \code{CHalfState} for the weights of \code{CMasslessHalfState} sorted 
      with respect to their transformation properties under the elements of the 
      centralizer,
\item \code{CState} for orbifold-invariant tensor products of massless left- 
      and right-movers, see Eq.~\eqref{eqn:invariantstate},
\item \code{COrbifold} for the full orbifold compactification, 
\item \code{CField} for a field of the effective 4-dimensional theory,
\item \code{CMonomial} for (gauge invariant) monomials of fields corresponding 
      to $D=0$ solutions,
\end{itemize}
and many more. In addition there are technical classes. For example, there are 
several classes devoted to group theory (like \code{dynkin}, \code{freudenthal}, 
\code{gaugeGroupFactor} and \code{gaugeGroup}), the class \code{CAnalyseModel} 
contains functions to analyze models for their phenomenological properties, the 
class \code{CPrint} contains all printing commands and the class \code{CPrompt} 
contains the source code of the \code{prompt}. As there are in total more than 
40 classes we cannot explain them in detail here. We give further details of 
some of the more important classes in~\ref{app:classes} and a short example 
program in~\ref{app:cppexample}.

%%%%%%%%%%%%%%%%%%% NEW SUBSECTION %%%%%%%%%%%%%%%%%%%

\subsection{Files defining an orbifold model}

There are two files that define an orbifold model: i) The {\it geometry file} 
contains, as the name suggests, the geometrical information about the orbifold, 
such as the space group, and ii) the {\it model file} contains shifts and 
Wilson lines, i.e. the action of the orbifold on the gauge sector of the 
heterotic string. In the following we give some more details. Examples are 
given in the additional material~\cite[\S Complementary notes]{ORB11}.

%%%%%%%%%%%%%%%%%%% NEW SUBSUBSECTION %%%%%%%%%%%%%%%%%%%

\subsubsection{The geometry file}
\label{sec:geometryfile}

The geometry file basically contains information about 
\begin{itemize}
\item the space group, i.e. the order of the twist(s), the six lattice vectors 
      of $\Gamma$ and the generators of the space group,
\item the discrete ($R$ and non-$R$) symmetries of the orbifold (important for 
      the computation of allowed superpotential couplings),
\item the inequivalent fixed points specified by their constructing elements 
      $\left( \theta^k\omega^l, n_\alpha e_\alpha\right)$ and
\item for each constructing element a list of centralizer elements, i.e. 
      elements $h \in S$ with $[h,g]=0$.
\end{itemize}
We give two examples in the additional material~\cite[\S Complementary notes]{ORB11}: 
Tab.~1 gives a detailed description of a geometry file using the \Z{3} example 
and Tab.~2 explains how to create a new geometry file of, for example, model 
(1-3) of Ref.~\cite{Donagi:2008xy}.

%%%%%%%%%%%%%%%%%%% NEW SUBSUBSECTION %%%%%%%%%%%%%%%%%%%

\subsubsection{The model file}
\label{sec:modelfile}

The model file contains a list of orbifold models, where each model is 
specified by
\begin{itemize}
\item the name of the model (will be used as the name of the corresponding 
      directory in the \code{prompt}),
\item the name of the geometry file,
\item the type of heterotic string (i.e. $\text{Spin}(32)/\Z{2}$ or 
      $\E{8}\x\E{8}$),
\item two shifts $V_1$, $V_2$ (where the $V_2 = 0$ for \Z{M} orbifolds),
\item six Wilson lines $W_\alpha$,
\item optionally, the parameters of (generalized) discrete torsion $a$, 
      $b_\alpha$, $c_\alpha$ and $d_{\alpha\beta}$ as defined in 
      Ref.~\cite{Ploger:2007iq},
\item optionally, some of the $\U{1}$ generators,
\item optionally, one can specify a script that is executed 
      automatically after the model has been loaded. 
\end{itemize}
An example for the case of a \Z{3} orbifold with standard embedding is given in 
the additional material~\cite[\S Complementary notes]{ORB11}.

%%%%%%%%%%%%%%%%%%% NEW SECTION %%%%%%%%%%%%%%%%%%%
\section{Conclusions and outlook}
\label{sec:conclusions}

With the tools provided here it should be possible to thoroughly investigate the
landscape (in particular the MSSM-like landscape) of orbifold
compactification of the heterotic string.  This ``heterotic
braneworld" provides a coherent geometric picture of MSSM-like
models. Crucial properties of the scheme depend on the geography of
fields in extra dimensions. This leads to the concept of ``Local Grand
Unification" and a geometric understanding of e.g. Yukawa couplings,
the $\mu$-term, neutrino masses and proton decay. Detailed properties
of models can be computed reliably within the context of conformal
field theory.

The models constructed here should be compared (and possibly related)
to other regions of the MSSM-like landscape, as e.g.~fermionic
formulations~\cite{Faraggi:1989ka,Faraggi:1992fa,Assel:2010wj},
tensoring of conformed field theories~\cite{Dijkstra:2004cc}, smooth
compactifications of the heterotic
string~\cite{Candelas:1985en,Donagi:2004ub,Braun:2005ux,Bouchard:2005ag,Anderson:2011cza,Anderson:2011ns},
type II (intersecting)
brane-models~\cite{Blumenhagen:2005mu,Gmeiner:2005vz,Blumenhagen:2006ci,Douglas:2006xy,Gmeiner:2008xq},
M- and F-theory
constructions~\cite{Vafa:1996xn,Donagi:1998xe,Donagi:1999ez,Beasley:2008dc,Beasley:2008kw,Acharya:2008zi,Weigand:2010wm}. It
would be interesting to identify similarities and differences of the
corresponding schemes. All these cases rely on a (sometimes hidden)
geometric interpretation that defines the properties of the models
such as the appearance of grand unification, values of
Yukawa-couplings and the existence of hierarchies.

One of the important observations in the framework of the heterotic braneworld 
concerns the crucial role played by discrete (gauge) symmetries.
They arise as remnants of the gauge symmetry as well
as symmetries due to the special location of fields
in extra dimensions. They 
control properties of the scheme, as e.g.~flavor universality and
the question of proton stability. At the orbifold point we encounter an
enhancement of discrete symmetries and particle spectra. These symmetries
are a basic ingredient of model building. Slightly broken, they might give
us an explanation for the appearance of hierarchies in particle
physics (as e.g. the ratio of Yukawa couplings). At the orbifold point we can 
rely on exact conformal field theory techniques that could be useful
to understand the blow-up procedure~\cite{Dixon:1986qv,Hamidi:1986vh,Aspinwall:1994ev,Lust:2006zh,Nibbelink:2007pn,Blaszczyk:2010db,Blaszczyk:2011ig}
of orbifold singularities in a 
controlled way and thus connect to smooth compactifications.

With a better knowledge of the MSSM-landscape we might hope to relate the
various constructions and improve the calculational power in those models
where we still have to rely on an effective low-energy supergravity and/or
large volume approximation. The future of the field requires reliable
calculational tools (as e.g. conformed field theory techniques) which are
up to now only applicable in some corners of the landscape. But we might
be lucky and nature might have chosen to live close to such a corner.

\section*{Acknowledgments}

We would like to thank Michael Blaszczyk, Nana Geraldine Cabo Bizet, Stefan 
F\"orste, David Grell\-scheid, Tom\'a\v{s} Je\v{z}o, Stefan Groot Nibbelink, 
Michael Ratz, Fabian R\"uhle and  Jes\'us Torrado-Cacho for useful discussions. 
P.V. would like to thank LMU Excellent for support. This 
material is partly based upon work done at the Aspen Center for Physics and 
supported by the National Science Foundation under Grant No. 1066293. This 
research was supported by the DFG cluster of excellence Origin and Structure of 
the Universe, the SFB-Transregio TR33 ``The Dark Universe" and TR27 
``Neutrinos and Beyond" (Deutsche Forschungsgemeinschaft) and the European 
Union 7th network program ``Unification in the LHC era" (PITN-GA-2009-237920). 
S.~R-S. was partially supported by CONACyT project 82291 and DGAPA project 
IA101811.

\appendix
\clearpage
\newpage

\section{Glossary of relevant classes}
\label{app:classes}

The \orbifolder{} makes extensive use of the class structure offered by 
\code{C++}. The information of an orbifold model is distributed according to the 
following classes.

\paragraph{Class \code{CAnalyseModel}} The class \code{CAnalyseModel} contains 
several functions that analyze the phenomenology of orbifold models, mainly for 
the cases of the Standard Model, Pati Salam or $\SU{5}$ gauge group.

\paragraph{Class \code{CField}} A \code{CField} object contains all physical 
information about a massless field, such as the representation, the \U{1} 
charges, the $q_\text{sh}$ charges, the localization and its vev. In addition, 
it contains a list of indices which specify its weights $p_\text{sh}$ (stored 
in the member variable \code{vector<CVector> Weights} of the associated 
\code{CMassless\-Half\-State} object).

\paragraph{Class \code{CFixedBrane}} A \code{CFixedBrane} object contains all 
information about all (untwisted or twisted) strings with constructing element 
$g = \left(\theta^k \omega^l ,n_\alpha e_\alpha\right) \in S$. The left-moving 
part of the string is computed using the local shift $V_g = k V_1 + l V_2 + 
n_\alpha W_\alpha$. Hence, the solutions of the equation for massless 
left-movers, Eq.~\eqref{eqn:massless}, are stored here in a vector of 
\code{CMasslessHalfState} objects, one entry for each different choice of 
oscillator excitation. After the massless solutions have been identified, they 
are sorted with respect to their centralizer eigenvalues and stored in a 
corresponding vector of \code{CHalfState} objects (one entry for each different 
choice of $\tilde{N}$ with different eigenvalues). In the last step, the 
massless right-moving \code{CHalfState} objects from \code{CSector} are 
tensored together with the massless left-moving \code{CHalfState} objects 
stored here to form orbifold-invariant string states. These states are stored 
in \code{vector<CState> InvariantStates}.

\paragraph{Class \code{CHalfState}} A \code{CHalfState} object descends from a 
\code{CMasslessHalfState} object by sorting the massless solutions (i.e. the 
weights $q_\text{sh}$ or $p_\text{sh}$ for right- or left- movers) with respect 
to their centralizer-eigenvalues. The indices of the weights (as listed in 
\code{vector<CVector> Weights} of the corresponding \code{CMasslessHalfState} 
object) having the same eigenvalues \code{vector<double> Eigenvalues} are 
stored in \code{vector<unsigned> Weights}.

\paragraph{Class \code{CMasslessHalfState}} A \code{CMasslessHalfState} object 
stores the solutions of the equation for massless right- or left-movers, 
respectively, see Eq.~\eqref{eqn:massless}. The constructor 
\code{CMasslessHalfState(MoversType Type, const S\_OscillatorExcitation 
\&Excitation)} needs two parameters: the first parameter \code{Movers\-Type} 
can be either \code{LeftMover} or \code{RightMover} and the second one 
specifies the oscillator excitation. Then, one can call the member function 
\code{bool SolveMassEquation(const CVector \&constructing\_Element, const 
SelfDualLattice \&Lattice)} to create the solutions of 
Eq.~\eqref{eqn:massless}, where \code{constructing\_Element} denotes the local 
twist $v_g$ or the local shift $V_g$ and \code{SelfDualLattice} can be either 
\code{E8xE8}, \code{Spin32} or \code{SO8}. The solutions are stored in 
\code{vector<CVector> Weights}.

\paragraph{Class \code{COrbifold}} A \code{COrbifold} object contains all 
information about a single orbifold compactification. The main member variable 
is \code{vector<CSector> Sectors}, i.e. a vector of $M$ times $N$ CSector 
objects, one for each sector of a \Z{M}\x\Z{N} orbifold. The first element 
corresponds to the untwisted sector and the rest to the various twisted sectors.

\paragraph{Class \code{COrbifoldGroup}} A \code{COrbifoldGroup} object 
basically contains the space group and its gauge embedding as objects of class 
\code{CSpaceGroup}, \code{CShiftVector} and \code{CWilsonLines}. Furthermore, 
it contains a vector of all inequivalent constructing elements 
(\code{vector<COrbifoldGroupElement>}) and a corresponding vector of 
centralizer elements (\code{vector<vector<COrbifoldGroupElement> >}).

\paragraph{Class \code{CPrint}} The \code{CPrint} class contains all printing 
commands. For the constructor \code{CPrint(OutputType output\_type, ostream 
*out)} one needs to specify the \code{OutputType}, being either 
\code{Tstandard}, \code{Tmathematica}, or \code{Tlatex}, and a \code{ostream} 
object to set the destination of the output, either to the screen using 
\code{\&cout} or to a file using an \code{ofstream} object.

\paragraph{Class \code{CSector}} A \code{CSector} object contains all 
information about an untwisted or twisted sector. It is mainly specified by the 
local twist $v_g$ (or in other words by $k$ and $l$ since $v_g = k v_1 + l 
v_2$). As the oscillator excitations and the right-moving part of the string 
only depend on the local twist, see Eq.~\eqref{eqn:massless}, they are 
identical for all strings from a given sector. Hence, this data is stored in 
\code{CSector}.

\paragraph{Class \code{CSpaceGroup}} A \code{CSpaceGroup} object 
contains the details about the space group of an orbifold model, as explained 
in Section~\ref{sec:orbis}. All constructing and centralizer elements are 
stored in \code{CSpaceGroup\-Element} objects. Additionally, it includes the 
geometrical information of the compact space, such as the 6D lattice, its 
symmetries and the order of the associated Wilson lines.

\paragraph{Class \code{CState}} A \code{CState} object is basically an 
orbifold-invariant combination of a massless right-moving \code{CHalfState} 
object and a massless left-moving \code{CHalfState} object.

\subsection{Example source code}
\label{app:cppexample}

We present a sample program that computes and analyzes the spectrum of the 
\Z{6}--II orbifold model of \cite{Kobayashi:2004ya}. In the source code 
distribution, the corresponding file is \code{src/examples/samplemain01.cpp}.

\begin{Verbatim}[fontsize=\small,numbers=left,xleftmargin=20pt,formatcom=\color{gray}]
#include <stdio.h>
#include "cprompt.h"
using namespace std;

int main(int argc, char *argv[])
{
  ifstream in("modelKRZ_A1.txt");
  if((!in.is_open()) || (!in.good()))
    exit(1);

  CPrint Print(Tstandard, &cout);
  string ProgramFilename = "";

  COrbifoldGroup OrbifoldGroup;
  if (OrbifoldGroup.LoadOrbifoldGroup(in, ProgramFilename))// load from file
  {
    cout << "\n-> Model file \"modelKRZ_A1.txt\" loaded." << endl;
    COrbifold KRZ_A1(OrbifoldGroup);                       // create the orbifold
    cout << "-> Orbifold \"KRZ_A1\" created.\n" << endl;

    cout << "-> Print shift and Wilson lines:" << endl;
    Print.PrintShift(OrbifoldGroup.GetShift(0));
    Print.PrintWilsonLines(OrbifoldGroup.GetWilsonLines(), true);

    cout << "\n-> Print spectrum, first with and then without U(1) charges:" << endl;
    Print.PrintSummaryOfVEVConfig(KRZ_A1.StandardConfig);  // print with U(1)s
    SConfig VEVConfig = KRZ_A1.StandardConfig;             // create new vev-config.
    VEVConfig.ConfigLabel = "TestConfig";                  // rename new vev-config.
    VEVConfig.SymmetryGroup.observable_sector_U1s.clear(); // change obs. sector
    Print.PrintSummaryOfVEVConfig(VEVConfig);              // print without U(1)s

    cout << "-> Analyze model:" << endl;
    vector<SConfig> AllVEVConfigs;
    bool SM = true;                                        // look for SM
    bool PS = true;                                        // look for PS models
    bool SU5 = true;                                       // look for SU(5) models
    // analyze the configuration "KRZ_A1.StandardConfig" of "KRZ_A1"
    // and save the result to "AllVEVConfigs"
    CAnalyseModel Analyze;
    Analyze.AnalyseModel(KRZ_A1, KRZ_A1.StandardConfig, SM, PS, SU5, 
    AllVEVConfigs, Print);
    if (SM || PS || SU5) // if one of the three possibilities is true 
    {
      cout << "-> Model has 3 generations plus vector-like exotics:" << endl;
      const size_t s1 = AllVEVConfigs.size();              // print all new configs.
      for (unsigned i = 0; i < s1; ++i)
        Print.PrintSummaryOfVEVConfig(AllVEVConfigs[i], LeftChiral, true);
    }
  }
  return EXIT_SUCCESS;
}
\end{Verbatim}
\label{verb:sampleprogram}

First, we define an \code{ifstream} object called \code{in} that contains the 
model file \code{modelKRZ\_A1.txt}. Then, we define the orbifold group as a 
\code{COrbifoldGroup} object and load the content of \code{in}. If no error 
occurs, the orbifold is defined as a \code{COrbifold} object. After calling the 
constructor of \code{COrbifold} with a \code{COrbifoldGroup} parameter, the 
spectrum of the orbifold model is computed and checked for consistency. 
Next, we print the shift, the Wilson lines and the massless spectrum of the 
model using the \code{CPrint} class. In the last part, a \code{CAnalyseModel} 
object is constructed in order to analyze the phenomenological properties of 
the vev-configuration \code{KRZ\_A1.StandardConfig}.

%%%%%%%%%%%%%%%%%%% NEW SECTION %%%%%%%%%%%%%%%%%%%

\section{Glossary of commands}
\label{app:glossarycommands}

In this appendix, we give short explanations for all commands of the 
\code{prompt} and of the web interface. In~\ref{app:conceptsandgeneralcommands} 
we start with some concepts and general commands. Then, in~\ref{app:directories} 
we list all commands available in the various directories of the \code{prompt}.

%%%%%%%%%%%%%%%%%%% NEW SUBSECTION %%%%%%%%%%%%%%%%%%%

\subsection{Concepts and general commands}
\label{app:conceptsandgeneralcommands}

The basic quantities of the \code{prompt} are fields of the 4D 
effective field theory (\ref{app:fieldlabels}). In order to access them easily 
one can define sets of fields (\ref{app:sets}). Furthermore, gauge invariant 
monomials of fields are used to describe solutions of the $D=0$ condition 
(\ref{app:monomials}). For many commands dealing with fields one can 
use the parameter \code{if(condition)} to choose only those fields that fulfill 
the condition (\ref{app:ifcondition}). Finally, this section describes the 
concept of processes (\ref{app:processes}), the use of vectors 
(\ref{app:vectors}), how to change the typesetting to \code{mathematica} or to 
\code{latex} style (\ref{app:outputstyle}), the use of system commands and 
variables (\ref{app:system}).

%%%%%%%%%%%%%%%%%%% NEW SUBSUBSECTION %%%%%%%%%%%%%%%%%%%

\subsubsection{Field labels}
\label{app:fieldlabels}

For a given orbifold model and vev-configuration, fields of the 4D effective 
field theory are tagged with labels, for example \code{q\_1}, \code{q\_2} and 
\code{q\_3} for the three left-handed quark doublets. A label consists of a 
name (\code{q}) and a generation index (\code{1}, \code{2} and \code{3} in our 
example). One can access several fields simultaneously using their common name, 
for example \code{q} for the three quarks. Furthermore, one can access all 
fields of a given model using \code{*}. In addition, one can obtain the 
intersection of all fields named \code{A} but not named \code{B} using 
\code{A-B}. Examples for intersections are: \code{n-n\_1} to get all fields 
named \code{n} except for \code{n\_1} and \code{*-n} for all fields except for 
the ones named \code{n}. 

Field labels are stored in the currently used vev-configuration of the orbifold 
model. They can be viewed and changed in the directory 
\code{/vev-config/labels>}, see~\ref{app:directorylabels}. Note that in a given 
vev-configuration one can define several labels for each field.

Finally, in \code{mathematica} typesetting, for example, the label \code{q\_1} 
is displayed as \code{fldq1}.

%%%%%%%%%%%%%%%%%%% NEW SUBSUBSECTION %%%%%%%%%%%%%%%%%%%

\subsubsection{Sets of fields}
\label{app:sets}

One can access several fields simultaneously not only by their field labels but 
also using sets of fields. These sets are stored in the currently used 
vev-configuration of the orbifold model. (Consequently, one cannot access a set 
in a different vev-configuration than in the one where it was created.) For 
more details on vacua, see~\ref{app:directoryconfig}. Note that sets are on the 
same footing as field labels\footnote{Both, fields and sets of fields, will be 
denoted as \code{fields} in the explanations of the following sections.}. 
I.e.~one can build intersections like: 
\begin{itemize}
\item \code{A-B} for the intersection of two sets \code{A} and \code{B}, 
\item \code{*-A} for the intersection of all fields \code{*} and a set \code{A}, 
\item \code{A-q} for the intersection of a set \code{A} and all fields of name 
      \code{q} or 
\item \code{q-A} for the intersection of all fields of name \code{q} and a set 
      \code{A}.
\end{itemize}

The commands to create and manipulate sets are displayed in any orbifold model 
directory of the prompt using the command 
\begin{equation}
\text{\code{help sets}}\;.
\end{equation}
The commands are:

\myvspace
\paragraph{Command \code{create set(SetLabel)}} Create an empty set with name 
\code{SetLabel} and save it in the currently used vev-configuration. Optionally, 
this command allows for the parameters \code{from monomials} or \code{from 
monomial(MonomialLabel)} in which case all fields from either all monomials or 
only from monomial \code{MonomialLabel} will be inserted into the new set. 
See~\ref{app:monomials} for more details on monomials.

\myvspace
\paragraph{Command \code{delete set(SetLabel)}} Delete the set \code{SetLabel} 
of the currently used vev-configuration.

\myvspace
\paragraph{Command \code{delete sets}} Delete all sets of the currently used 
vev-configuration.

\myvspace
\paragraph{Command \code{insert(fields) into set(SetLabel)}} Insert 
\code{fields} into the set \code{SetLabel}. Optionally, the parameter 
\code{if(condition)} can be used to insert only those \code{fields} into the 
set \code{SetLabel} that satisfy the condition. For details on 
\code{if(condition)} see~\ref{app:ifcondition}.

\myvspace
\paragraph{Command \code{print set(SetLabel)}} Print the content of the set 
\code{SetLabel}.

\myvspace
\paragraph{Command \code{print sets}} Print all sets defined in the currently 
used vev-configuration. One can use the optional parameter \code{if not empty} 
to print only the non-empty sets.

\myvspace
\paragraph{Command \code{remove(fields) from set(SetLabel)}} Remove 
\code{fields} from the set \code{SetLabel}. Optionally, the parameter 
\code{if(condition)} can be used to remove only those fields that satisfy the 
condition.

\myvspace
\paragraph{Command \code{\#fields in set(SetLabel)}} Count the number of fields 
in the set \code{SetLabel}.

A short example showing some of the basic commands for sets is given in 
Tab.~\ref{tab:examplesets}. 

\begin{table}[t]
\begin{center}
\begin{tabular}{|@{\;\small\tt}l|@{\normalsize}|l|}
\hline
command                                         & description\\
\hline
\hline
create set(Test)                                & create an empty set named 
                                                  \code{Test} \\
insert(F\_1 F\_2 F\_3 F\_4 F\_5) into set(Test) & insert \code{F\_i}, \code{i} 
                                                  $=1,\ldots,5$ into the set 
                                                  \code{Test} \\
remove(*) from set(Test) if(\#osci. != 0)       & remove fields with non-zero 
                                                  number operator $\tilde{N}$ 
                                                  (i.e. \code{F\_5})\\
print sets                                      & print all sets \\
\hline
\end{tabular}
\end{center}
\caption{Short example for the use of set-commands in the directory 
{\code{/Z3StandardEmbedding>}} of the \Z{3} standard embedding model 
(using the standard labels {\code{F\_i}} of the vev-configuration 
\code{TestConfig1}).}
\label{tab:examplesets}
\end{table}

%%%%%%%%%%%%%%%%%%% NEW SUBSUBSECTION %%%%%%%%%%%%%%%%%%%

\subsubsection{Gauge invariant monomials}
\label{app:monomials}

(Holomorphic) gauge invariant monomials (short: monomials) are used to describe 
solutions to the $D=0$ supersymmetry condition 
\cite{Buccella:1982nx, Cleaver:1997jb, Kappl:2011vi, Dcode:2011df}. A (sub-)set 
of solutions can be found using the command \code{find D-flat(fields)} 
described in \ref{app:directoryconfig}. More details and examples can be seen 
using the command
\begin{equation}
\text{\code{help monomials}}
\end{equation}
in any orbifold model directory.

%%%%%%%%%%%%%%%%%%% NEW SUBSUBSECTION %%%%%%%%%%%%%%%%%%%

\subsubsection{If conditions}
\label{app:ifcondition}

Many commands that deal with fields allow for the parameter 
\code{if(condition)} (or several copies thereof) so that only those fields are 
chosen that fulfill all the conditions. An explicit example was already given 
in Tab.~\ref{tab:examplesets}. In general, a condition consists of three parts:
\begin{itemize}
\item the left hand side gives the variable (e.g. \code{Q\_i} for the fields 
      \code{i}-th $\U{1}$ charge, \code{vev} for the fields vacuum expectation 
       value or \code{\#osci.} for the number of oscillators),
\item the middle gives the comparison operator (e.g. \code{==} for equal or 
      \code{!=} for unequal) and 
\item the right hand side gives a value (e.g. a rational number or $0$).
\end{itemize}
More details and examples can be seen using the command
\begin{equation}
\text{\code{help conditions}}
\end{equation}
in any orbifold model directory.

%%%%%%%%%%%%%%%%%%% NEW SUBSUBSECTION %%%%%%%%%%%%%%%%%%%

\subsubsection{Processes}
\label{app:processes}
The following commands start new processes that run in the background so that 
one can continue to work with the \code{prompt}:
\begin{itemize}
\item \code{/> create random orbifold from(OrbifoldLabel)}
\item \code{/A/couplings> create coupling(fields)}
\item \code{/A/vev-config> find D-flat(fields)}
\item \code{/A/gauge group> find accidental U1s}
\end{itemize}
Each process has an ID, the so called PID. Similar to the Linux command line 
one can see all running processes using the command \code{ps} and terminate a 
process with PID \code{A} using \code{kill(A)}. One can also kill all active 
processes using the command \code{kill(all)}. In a script the command 
\code{wait(X)} might be useful in order to check every \code{X} seconds if all 
processes have finished and to continue with the next commands afterwards. More 
details can be seen using the command
\begin{equation}
\text{\code{help processes}}
\end{equation}
in any orbifold model directory.

%%%%%%%%%%%%%%%%%%% NEW SUBSUBSECTION %%%%%%%%%%%%%%%%%%%

\subsubsection{Vectors}
\label{app:vectors}
Many commands need a vector of rational numbers as a parameter. Examples 
include the commands \code{set shift V(i) = <16D vector>} and \code{set torsion 
b = <6D vector>}. In these cases there are several possibilities of how to 
write the vector. For example, the following forms of a \code{<4D vector>} are 
possible:
\begin{equation}
(1/3\; 1/1\; 0/1\; 0/1)\;=\;(1/3\; 1\; 0\; 0)\;=\;(1/3,1,0,0)\;=\;(1/3,1,0\text{\^{}}2)\;=\;1/3(1\; 3\; 0\text{\^{}}2)
\end{equation}
In addition, for the first four forms of the example-vector, one can leave the 
brackets away.

%%%%%%%%%%%%%%%%%%% NEW SUBSUBSECTION %%%%%%%%%%%%%%%%%%%

\subsubsection{Output for mathematica or in latex style}
\label{app:outputstyle}

Often it is useful to transfer data from the \orbifolder{} to mathematica, 
for example, in order to use \code{STRINGVACUA} \cite{Gray:2008zs}, 
\code{SINGULAR} \cite{DGPS}, \code{NonAbelianHilbert} \cite{Kappl:2011vi, 
Dcode:2011df} or \code{DiscreteBreaking} \cite{Petersen:2009ip, ZNcode:2009zn}. 
Therefore, many commands allow for the parameter
\begin{equation}
\text{\code{@mathematica}}
\end{equation}
so that the output of the command will be printed in a mathematica compatible 
style (if available). For example, 
\begin{equation}
\text{\code{print list of charges @mathematica}}
\end{equation}
in the directory \code{/spectrum>}. Similarly, the parameter \code{@latex} can 
be used in order to get the output in latex code (again, if available). In 
addition, one can set the default typesetting to \code{mathematica}, 
\code{latex} or back to \code{standard} using the commands
\begin{equation}
\text{\code{@typesetting(mathematica)}},\; \text{\code{@typesetting(latex)}} 
\quad\text{or}\quad \text{\code{@typesetting(standard)}}\;,
\end{equation}
respectively. Finally, the parameter \code{no output} can be used to suppress 
the output of the current command.

%%%%%%%%%%%%%%%%%%% NEW SUBSUBSECTION %%%%%%%%%%%%%%%%%%%

\subsubsection{System commands and variables}
\label{app:system}

System commands start with the symbol \code{@} and are used to change the 
output's style and destination. Moreover, the \code{prompt} allows for some 
pre-defined variables which are particularly usefull in scripts. They 
start and end with the symbol \code{\$}.

\myvspace
\paragraph{Command \code{@typesetting(Type)}} Change the output's typesetting, 
see~\ref{app:outputstyle}.

\myvspace
\paragraph{Command \code{@begin print to file(A)}} Start printing output to 
file \code{A} and not to the screen. In contrast, one can use the parameter 
\code{to file(A)} so that the output of only the current command is 
printed to file, e.g. \code{print summary to file(Summary.txt)}.

\myvspace
\paragraph{Command \code{@end print to file}} Stop printing output to file.

\myvspace
\paragraph{Command \code{@status}} Display the destination of the output (e.g. 
\code{screen}) and the style of the typesetting (i.e. \code{standard}, 
\code{latex} or \code{mathematica}).

\myvspace
\paragraph{Variables} There are three pre-defined variables:
 \code{\$OrbifoldLabel\$}, \code{\$VEVConfigLabel\$} and \code{\$Direc\-tory\$}. 
When executed, a variable is replaced by a corresponding string, being the 
label of the current orbifold model, the label of the current vev-configuration 
or the path of the current directory, respectively. They are particularly 
usefull in scripts, e.g. used as \code{to file(\$OrbifoldLabel\$.txt)}.

%%%%%%%%%%%%%%%%%%% NEW SUBSECTION %%%%%%%%%%%%%%%%%%%

\subsection{The directories}
\label{app:directories}

The structure of the \code{prompt} consists of a main directory \code{/>} and 
subdirectories that correspond to orbifold models. Each orbifold model 
directory has further subdirectories \code{/model>}, \code{/gauge group>}, 
\code{/spectrum>}, \code{/couplings>} and \code{/vev-config>}. They offer 
commands of the respective category. In this section we give an alphabetically 
ordered glossary of directory-commands and explain their use in detail.

\subsubsection{The main directory \code{/>}}
\label{app:maindirectory}

In the main directory one can basically create, load and save orbifold models.

\myvspace
\paragraph{Command \code{create orbifold(OrbifoldLabel) with point group(M,N)}} 
Create an empty orbifold model directory for an orbifold of specified point 
group orders (use \code{N}$=1$ for \Z{M} orbifolds).

\myvspace
\paragraph{Command \code{create random orbifold from(OrbifoldLabel)}} Randomly 
create new orbifold models. Details are given in Section 
\ref{sec:createneworbifolds}. More details and examples can be seen using the 
main directory's \code{/>} command
\begin{equation}
\text{\code{help create random}}\;.
\end{equation}

\myvspace
\paragraph{Command \code{delete orbifold(OrbifoldLabel)}} Delete the orbifold 
model directory \code{OrbifoldLabel}.

\myvspace
\paragraph{Command \code{delete orbifolds}} Delete all orbifold model 
directories.

\myvspace
\paragraph{Command \code{load orbifolds(Filename)}} Load all orbifold models 
from the model file named \code{Filename}.

\myvspace
\paragraph{Command \code{load program(Filename)}} Load a script from file 
\code{Filename} and execute the commands contained in that file.

\myvspace
\paragraph{Command \code{save orbifolds(Filename)}} Save all orbifold models of 
the main directory to a model file named \code{Filename}.

\subsubsection{The directory \code{/model>}}
\label{app:directorymodel}

In the directory \code{/model>} the input data (e.g. point group, twists, 
shifts, Wilson lines, etc.) of the current orbifold model can be displayed and 
changed. 

\myvspace
\paragraph{Command \code{create suborbifold with factor(i)}} Starting from an 
orbifold with space group $S$, one can create a so-called suborbifold based on 
a subgroup $S' \subset S$. 

For \Z{M} orbifolds the subgroup $S' \subset S$ is specified by one number, $i$ 
being a divisor of $M$. Denote the \Z{M} twist generator of the space group by 
$g \in S$. Then, the subgroup $S' \subset S$ is based on the twist generator 
$g^i \in S' \subset S$ which generates \Z{M/i}. One can use the optional 
parameter \code{and(j)} to choose $g^i \in S' \subset S$ and $g^j \in S' 
\subset S$ (with $i$, $j$ coprime and $i$, $j$ divide $M$) to generator 
\Z{M/i}\x\Z{M/j}.

In the case of \Z{M}\x\Z{N} orbifolds (with twist generators $g_1, g_2 \in S$) 
one has to specify two numbers \code{create suborbifold with factor(i,j)} 
(where $i$ divides 
$M$ and $j$ divides $N$) for the new generator $g_1^i g_2^j$ of the subgroup 
$S' \subset S$. Again, one can use the optional parameter \code{and(k,l)} to 
specify a second generator $g_1^k g_2^l$.

Note that this command is particularly useful to analyze the 6D orbifold GUT 
limit of an orbifold model. For example, start with the \Z{6}--II orbifold MSSM 
of \cite{Buchmuller:2006ik}. Then the command \code{create suborbifold with 
factor(2)} will produce the 6D \Z{3} orbifold GUT limit as analyzed in 
\cite{Buchmuller:2007qf}.

\myvspace
\paragraph{Command \code{print available space groups}} Print a list of all 
geometry files compatible with the specified point group. The geometry files 
are searched by the \orbifolder{} in the directory \code{/localdirectory/} 
\; \code{Geometry>} (of the local PC). For more details on the content of 
geometry files, see Section \ref{sec:geometryfile}.

\myvspace
\paragraph{Command \code{print discrete symmetries}} Print the discrete ($R$ 
and non-$R$) symmetries as defined in the geometry file. Note that $R$ 
symmetries need to be defined in the geometry file in order to be used in the 
computation of allowed superpotential couplings.

\myvspace
\paragraph{Command \code{print discrete torsion}} Print the (generalized) 
discrete torsion parameters $a$, $b_\alpha$, $c_\alpha$ and $d_{\alpha\beta}$ 
as defined in Ref.~\cite{Ploger:2007iq}.

\myvspace
\paragraph{Command \code{print massless A-movers}} where \code{A} can be 
\code{left} or \code{right}. Print the massless left- or right movers before 
some of them are projected out by the action of the centralizer.

\myvspace
\paragraph{Command \code{print orbifold label}} Print the orbifold label (i.e. 
the name of the current orbifold directory).

\myvspace
\paragraph{Command \code{print point group}} Print the point group. The output 
reads, for example, \code{  Point group is Z\_3.}

\myvspace
\paragraph{Command \code{print shift}} Print the shift(s) as 16D vector(s). 
The output reads e.g.
\begin{equation}
\text{\code{  V\_1 = (  1/3   1/3  -2/3     0     0     0     0     0) 
                     (    0     0     0     0     0     0     0     0) }}\;.
\end{equation}

\myvspace
\paragraph{Command \code{print space group}} First, print the point group and 
the root-lattice. Next, print the generators of the space group. The output 
reads e.g.
\begin{eqnarray}
&&\text{\code{  Space group based on Z\_3 point group and root-lattice of SU(3)\^{}3.}} \\
&&\text{\code{  Generators are:}}\nonumber \\
&&\text{\code{  (1, 0) (    0,     0,     0,     0,     0,     0) }} \nonumber \\
&&\text{\code{  (0, 0) (    1,     0,     0,     0,     0,     0) }} \cdots \nonumber
\end{eqnarray}
where \code{(k,l) (n\_1, \ldots, n\_6)} corresponds to the element 
$\left(\theta^k\omega^l, n_\alpha e_\alpha\right)$ of the space group. Note 
that roto-translations and freely-acting involutions are allowed as generators 
of the space group. For example in Ref.~\cite{Donagi:2008xy}, one of the 
generators of the (0-2) model reads \code{(0,1) (0,0,0,0,1/2,0)} corresponding 
to $\left(\omega, \frac{1}{2}e_5\right)$ and one of the generators of the (1-1) 
model reads \code{(0,0) (0,1/2,0,1/2,0,1/2)} corresponding to $\left(\Id, 
\frac{1}{2}(e_2+e_4+e_6)\right)$.

\myvspace
\paragraph{Command \code{print twist}} Print the twist(s) as four-dimensional 
vector(s). The output reads e.g.
\begin{equation}
\text{\code{  v\_1 = (    0   1/3   1/3  -2/3) }}\;.
\end{equation}

\myvspace
\paragraph{Command \code{print Wilson lines}} Print the relations among the 
Wilson lines (e.g. $W_1 = W_2$ for \Z{3}), their order (e.g. order 3 for $3W_i 
\in \Lambda$ for \Z{3}) and the Wilson lines themselves as 16D vectors. The 
output reads, for example,
\begin{eqnarray}
&&\text{\code{  Wilson lines identified on the orbifold: }} \\
&&\text{\code{    W\_1 = W\_2, W\_3 = W\_4, W\_5 = W\_6}} \nonumber \\
&&\text{\code{  Allowed orders of the Wilson lines: 3 3 3 3 3 3 }} \nonumber \\
&&\text{\code{  W\_1 = (    0     0     0     0     0     0     0     0) 
                (    0     0     0     0     0     0     0     0)}} \nonumber \\
&&\cdots\nonumber
\end{eqnarray}

\myvspace
\paragraph{Command \code{print \#SUSY}} Print the number of supersymmetry in 
4D. The output reads, for example,
\begin{equation}
\text{\code{N = 1 SUSY in 4D.}}
\end{equation}

\myvspace
\paragraph{Command \code{set heterotic string type(type)}} Define the 16D 
gauge lattice of the heterotic orbifold model. Here, \code{type} can be 
\code{E8xE8} or \code{Spin32}.

\myvspace
\paragraph{Command \code{set shift standard embedding}} Choose $V = (v^1, v^2, 
v^3, 0^{13})$ for \Z{M} orbifolds or $V_1 = $ \;\; $(v^1_1, v^2_1, v^3_1, 0^{13})$, 
$V_2 = (v^1_2, v^2_2, v^3_2, 0^{13})$ for \Z{M}\x\Z{N} orbifolds. (The notation
$0^{13}$ means 13 times the entry ``$0$".)

\myvspace
\paragraph{Command \code{set shift V = <16D vector>} or \code{set shift V(i) = 
<16D vector>}} Define the shift vector $V$ of \Z{M} orbifold models or one of 
the two shift vectors $V_i$ (with \code{i}$=1,2$) of \Z{M}\x\Z{N} orbifold 
models as 16D vector, see Eq.~\eqref{eq:gaugeembedding}. For more details 
on vectors see~\ref{app:vectors}.

\myvspace
\paragraph{Command \code{set torsion a = n/d}, \code{b = <6D vector>}, \code{c 
= <6D vector>} or \code{d = <15D vector>}} \quad Set the (generalized) discrete 
torsion parameters as defined in \cite{Ploger:2007iq}, i.e. $a$, $b_\alpha$, 
$c_\alpha$ and $d_{\alpha\beta}$ (for $\alpha, \beta = 1,\ldots, 6$; 
$d_{\alpha\beta} = -d_{\beta\alpha}$ has 15 components). Note that the 
parameters are not checked for modular invariance and hence might cause 
inconsistent spectra. For more details on vectors see~\ref{app:vectors}.

\myvspace
\paragraph{Command \code{set WL W(i) = <16D vector>}} Define the Wilson line 
$W_i$ as a 16D vector (with \code{i}$=1,\ldots,6$), see 
Eq.~\eqref{eq:gaugeembedding}. For more details on vectors 
see~\ref{app:vectors}.

\myvspace
\paragraph{Command \code{use space group(i)}} with \code{i}$ = 1, \ldots$. Load 
the space group from the \code{i}-th geometry file, where the index \code{i} 
corresponds to the position in the list of geometry files as displayed using 
the command \code{print available space groups}.

%%%%%%%%%%%%%%%%%%% NEW SUBSUBSECTION %%%%%%%%%%%%%%%%%%%

\subsubsection{The directory \code{/gauge group>}}
\label{app:directorygaugegroup}

In this directory one can print and change some details of the gauge group for 
the currently used vev-configuration. In more detail, one can display the 
$\U{1}$ generators and the simple roots, change the basis of \U1 generators, 
define a B-L generator and identify accidental \U1 symmetries of the 
superpotential. Note that all gauge-group-editing commands are not available in 
the vev-configuration \code{StandardConfig1}.

\myvspace
\paragraph{Command \code{delete accidental U1s}} Delete the accidental \U1 
charges of all fields.

\myvspace
\paragraph{Command \code{find accidental U1s}} Take the superpotential (as far 
as it has been created in the directory \code{/couplings>}, 
see~\ref{app:directorycouplings}) and identify its accidental $\U{1}$ 
symmetries. The command starts a new process that runs in the back of the 
\code{prompt}. The results are saved in the currently used vev-configuration. 
Presumably, the accidental $\U{1}$ symmetries will be broken explicitly by 
higher order terms, but nevertheless might be of phenomenological relevance, 
e.g. for the strong CP problem \cite{Choi:2009jt} and proton decay 
\cite{Forste:2010pf}.

Optionally, one can use the parameter \code{fields with zero charge(fields)} in 
order to find only those accidental $\U{1}$ symmetries under which the fields 
of \code{fields} are uncharged. On a technical level, this is achieved by 
inserting, during this analysis, each field of \code{fields} as a linear term 
into the superpotential.

\myvspace
\paragraph{Command \code{load accidental U1s(Filename)}} Load accidental 
$\U{1}$ charges from a file named \code{Filename}.

\myvspace
\paragraph{Command \code{print anomalous space group element}} Print details on 
discrete anomalies using the discrete symmetries defined in the geometry file 
and identify the so-called anomalous space group element \cite{Araki:2008ek}.

\myvspace
\paragraph{Command \code{print anomaly info}} Print details on gauge and 
gravitational anomalies and check their universality relations. In detail, in 
the case of $\mathcal{N} = 1$ SUSY in 4D, beside the pure non-Abelian 
anomalies, the relations\footnote{In the \orbifolder{}, the convention for the 
quadratic Dynkin index is such that $\ell=1$ for the fundamental representation 
of \SU{N} groups.}
\begin{equation}
\label{eq:Anomaly_conditions}
   \frac{1}{24}\text{tr}\,Q_i
 = \frac{1}{6 |t_i|^2}\text{tr}\,Q_i^3
 = \frac12\text{tr}\,\ell Q_i
 = \frac{1}{2|t_j|^2}\text{tr}\,Q_j^2 Q_i =
   \left\{ 
    \begin{array}{ll}
        \text{const.} \neq 0 \quad & \text{if } i = 1,\text{i.e. $i =$ anom} \\ 
        0 &   \text{otherwise}
    \end{array} 
   \right.
\end{equation}
(with $i\neq j$) are verified, where $t_i$ is a 16D vector corresponding to 
the $i$-th $\U{1}$ generator so that a field with shifted left-moving momentum 
$p_\text{sh}$ carries the charge $Q_i = p_\text{sh} \cdot t_i$ and $\text{tr}$ 
sums over the contributions from all massless left-chiral fields.

\myvspace
\paragraph{Command \code{print B-L generator}} Print the $\U1_{B-L}$ generator 
as a 16D vector.

\myvspace
\paragraph{Command \code{print FI term}} Print the Fayet-Iliopoulos term (i.e. 
$\text{tr}Q_\text{anom}$ as in Eq.~\eqref{eq:Anomaly_conditions}), if there is 
an anomalous $\U{1}$.

\myvspace
\paragraph{Command \code{print gauge group}} Print the observable and hidden 
part of the gauge group for the currently used vev-configuration.

\myvspace
\paragraph{Command \code{print simple root(i)}} Print the \code{i}-th simple 
root as 16D vector.

\myvspace
\paragraph{Command \code{print simple roots}} Print (a choice of) simple roots 
of all non-Abelian gauge group factors as 16D vectors.

\myvspace
\paragraph{Command \code{print U1 generator(i)}} Print the \code{i}-th $\U{1}$ 
generator as 16D vector.

\myvspace
\paragraph{Command \code{print U1 generators}} Print all $\U{1}$ generators as 
16D vectors.

\myvspace
\paragraph{Command \code{save accidental U1s(Filename)}} Save the accidental 
$\U{1}$ charges to a file named \code{Filename}.

\myvspace
\paragraph{Command \code{set B-L = <16D vector>}} Define $\U1_{B-L}$ as a 
16D vector. Since in the \code{or\-bi\-fol\-der} all \U1 generators are 
demanded to be orthogonal to each other, but $\U1_{B-L}$ is in general not 
orthogonal to hypercharge, B-L is stored as an additional vector. 
One can use the optional parameter \code{allow for anomalous B-L} if 
$\U1_{B-L}$ is allowed to mix with the anomalous \U1. For more details on 
vectors see~\ref{app:vectors}.

\myvspace
\paragraph{Command \code{set U1(i) = <16D vector>}} Change the basis of \U1 
generators by specifying a 16D vector as the $i$-th generator. The new 
generator must be orthogonal to all simple roots and to the $j$-th \U1 
generator, for $j < i$. Note that the $k$-th \U1 generators with $k>i$ will be 
changed automatically such that, at the end, all generators are orthogonal to 
each other. For more details on vectors see~\ref{app:vectors}.

%%%%%%%%%%%%%%%%%%% NEW SUBSUBSECTION %%%%%%%%%%%%%%%%%%%

\subsubsection{The directory \code{/spectrum>}}
\label{app:directoryspectrum}

This directory offers access to all information about the massless spectrum. In 
detail, for each massless field one can obtain the SUSY multiplet type (i.e. 
for $\mathcal{N} = 1$ supersymmetry in 4D: \code{left-chiral}, 
\code{right-chiral}, \code{vector} and \code{modulus}), the localization 
(corresponding to its constructing space group element), the shifted 
left-moving momentum $p_\text{sh}$, the non-Abelian representation, the \U{1} 
charges, the B-L charge (if defined), the shifted right-moving momentum 
$q_\text{sh}$, the oscillator excitations, the $R$ charges, modular weights (if defined), the 
label of the field and finally its vev. Note that complex conjugate 
representations are printed as negative integers; for example, \code{-3} 
denotes the conjugate fundamental representation \bsb{3} of \SU3.

\myvspace
\paragraph{Command \code{find potential blowup modes(fields)}} Print a list of 
potential blow-up modes considering \code{fields} only, i.e. print all 
\code{fields} for each fixed brane/point.

\myvspace
\paragraph{Command \code{find random blowup modes(fields)}} print a random list 
of blow-up modes \code{fields}, one per fixed brane/point. The result can be 
saved to a set \code{SetLabel} using the optional parameter \code{save to} \; 
\code{set(SetLabel)}.

\myvspace
\paragraph{Command \code{print(fields)}} Print some details of \code{fields}. 
There is one optional parameter, \code{with internal information}, that 
displays some internal information about how the fields' data can be accessed 
in the \code{C++} source code of the \orbifolder{}.

\myvspace
\paragraph{Command \code{print all states}} Print details of all fields 
(including left-chiral superfields, vector superfields of the (non-Abelian) 
gauge bosons, moduli and their CPT-partners).

\myvspace
\paragraph{Command \code{print list of empty fixed branes}} Print a list of all 
fixed branes/points that potentially could carry left-chiral fields but are 
empty for the current orbifold model.

\myvspace\vspace{-0.1cm}
\paragraph{Command \code{print list of charges(fields)}} Print gauge and 
discrete charges of \code{fields}. For example,
\begin{eqnarray}
  &\mbox{\footnotesize{\tt( -1/3  -1/3   2/3     0     0     0     0     0) 
  (    0     0     0     0     0     0     0     0)  
  ( -1/3   2/3  -1/3)  (    2     0     0     1)  "F\_10"}}& \\
  &\mbox{\footnotesize{\tt( -1/3   2/3  -1/3     0     0     0     0     0) 
  (    0     0     0     0     0     0     0     0)  
  ( -1/3   2/3  -1/3)  (    2     0     0     1)  "F\_10"}}& \nonumber\\
  &\mbox{\footnotesize{\tt(  2/3  -1/3  -1/3     0     0     0     0     0) 
  (    0     0     0     0     0     0     0     0)  
  ( -1/3   2/3  -1/3)  (    2     0     0     1)  "F\_10"}}& \nonumber
\end{eqnarray}
for the field \code{F\_10} of the \Z{3} orbifold with standard embedding. Each 
line consists of three parts:
\begin{itemize}
\item first the 16D shifted left-moving momentum $p_\text{sh}$ (printed as 
      two eight-dimensional vectors in the case of $\E{8}\x\E{8}$), 
\item the $R$ charges: $(R_1, R_2, R_3)=(-\tfrac{1}{3}, \tfrac{2}{3},
      -\tfrac{1}{3})$ in the example,
\item the charges with respect to the discrete non-$R$ symmetries: $(k, 
      n_1+n_2, n_3+n_4, n_5+n_6,)=(2, 0, 0, 1)$ in the example,
\item the label of the corresponding field,
\end{itemize}
where, in general, the $R$ and non-$R$ symmetries must be specified in the 
geometry file, see Section \ref{sec:geometryfile}. Note that it can be very 
helpful to use the optional parameter \code{@mathematica} for this command in 
order to transfer the information about the fields to mathematica, 
see~\ref{app:outputstyle}.

\myvspace
\paragraph{Command \code{print summary}} Print the gauge group and a summary 
table of the massless matter fields. Important optional parameters are:
\begin{itemize}
\item \code{of sectors}
\item \code{of fixed points}
\item \code{of fixed point(X)}\\ where the fixed point \code{X} can be 
      specified in three ways: i) using \code{k,l,n1,n2,n3,n4,n5,n6}, ii) using 
      \code{loc of F\_i} where \code{F\_i} is the label of a twisted field and 
      iii) by a fixed point label as specified in the directory 
      \code{/vev-config/labels>}.
\item \code{of sector T(k,l)} where k and l label the $\theta^\text{\code{k}} 
      \omega^\text{\code{l}}$ twisted sector. Use \code{of sector T(0,0)} for 
      the untwisted sector.
\end{itemize}
In all these cases one can use in addition the following, optional parameters:
\begin{itemize}
\item \code{with labels}: print the currently used labels of the fields as 
      specified in the directory \code{/vev-config/} \; \code{labels>}.
\item \code{no U1s}: do not print the $\U{1}$ charges of the fields.
\item type of SUSY multiplet, e.g. \code{vector}, \code{gravity}, \code{modulus} 
      or \code{anykind} for all types; if not specified the left-chiral fields 
      are printed only.
\end{itemize}
More details and examples can be seen using the command
\begin{equation}
\text{\code{help print summary}}
\end{equation}
in the directory \code{/spectrum>}.

\myvspace
\paragraph{Command \code{tex table(fields)}} Print a latex table with 
information about \code{fields}. The table contains the (gauge) charges with 
respect to the observable sector of the currently used vev-configuration and 
the discrete charges as specified in the geometry file (see Section 
\ref{sec:geometryfile}). One can use the optional parameter 
\code{print labels(i,j,..)} in order to list the \code{i}-th, \code{j}-th 
... label(s) of the fields.

%%%%%%%%%%%%%%%%%%% NEW SUBSUBSECTION %%%%%%%%%%%%%%%%%%%

\subsubsection{The directory \code{/couplings>}}
\label{app:directorycouplings}

The directory \code{/couplings>} allows to identify and analyze allowed terms 
of the superpotential (i.e. terms that are invariant under all gauge and 
discrete symmetries). The (in general moduli-dependent) coefficients are not 
computed. Furthermore, one can analyze mass matrices (e.g. of vector-like 
exotics).

Note that couplings and mass matrices are stored in the currently used 
vev-configuration. Hence, they can only be accessed in the vev-configuration 
where they have been defined. For simplicity, the abbreviations \code{mm} for 
\code{mass matrix} and \code{mms} for \code{mass matrices} apply to all 
commands.

\myvspace
\paragraph{Command \code{auto create mass matrix(A B)}} Create the couplings 
relevant for the effective mass matrix $M_{ij} A_i B_j$. Optionally, one can specify the 
the label of those fields whose vevs generate $M_{ij}$ using the parameter 
\code{singlet(N)} (with default value \code{N}=\code{n}) and the maximal order 
\code{X} in singlets \code{N} using the parameter \code{max order(X)}.

\myvspace
\paragraph{Command \code{create coupling(fields)}} Find the allowed 
superpotential-couplings between \code{fields} and store the result in the 
currently used vev-configuration. For example, all trilinear couplings are 
created using the command
\begin{equation}
\text{\code{create coupling(* * *)}}\;.
\end{equation}
Optionally, one can restrict \code{fields} using the parameter \code{allowed 
fields(...)}, e.g. the command \code{create coupling(n n n) allowed 
fields(SetA)} creates all trilinear couplings of fields \code{n} from 
\code{SetA}.

\myvspace
\paragraph{Command \code{find(fields)}} Displays a list of allowed couplings 
involving the fields \code{fields}.

\myvspace
\paragraph{Command \code{find effective(fields)}} As \code{find(fields)}, but 
only the effective couplings, i.e after replacing fields with non-zero vev by 
their vevs.

\myvspace
\paragraph{Command \code{load couplings(Filename)}} Load couplings from file 
\code{Filename} into the currently used vev-configuration. This command is 
disabled in the web interface.

\myvspace
\paragraph{Command \code{mass matrix(A B)}} Create the mass matrix $M_{ij} A_i 
B_j$ (from the current superpotential) and save it in the currently used 
vev-configuration.

\myvspace
\paragraph{Command \code{print effective superpotential}} Similar to the 
command \code{print superpotential} but print only the effective couplings, i.e 
after replacing fields with non-zero vev by their vevs.

\myvspace
\paragraph{Command \code{print list of mass matrices}} Print all mass matrices 
of the currently used vev-configuration.

\myvspace
\paragraph{Command \code{print mass matrix(i)}} Print the \code{i}-th mass 
matrix. The optional parameter \code{max order(X)} specifies the order \code{X} 
in the fields up to which a matrix entry shall be printed explicitly.

\myvspace
\paragraph{Command \code{print superpotential}} Print the superpotential of the 
currently used vev-configuration.

\myvspace
\paragraph{Command \code{print vanishing couplings}} Print all cases of 
vanishing couplings as lists of highest weights in Dynkin labels. See the 
command \code{remove vanishing couplings}.

\myvspace
\paragraph{Command \code{remove vanishing couplings}} Remove couplings that 
vanish because of symmetry/anti-sym\-metry of repeated identical fields, e.g. 
let $\ell$ be an $\SU{2}$ doublet, then the gauge invariant coupling $\ell 
\ell = \ell_i \ell_j \epsilon^{ij} = 0$ vanishes. This command requires 
additional user input. 

\myvspace
\paragraph{Command \code{save couplings(Filename)}} Save all couplings of the 
currently used vev-configuration to a file. One can optionally save only 
couplings of order~\code{X} using the parameter \code{of order(X)}. This 
command is disabled in the web interface.

%%%%%%%%%%%%%%%%%%% NEW SUBSUBSECTION %%%%%%%%%%%%%%%%%%%

\subsubsection{The directory \code{/vev-config>}}
\label{app:directoryconfig}

In this directory one can define several vev-configurations. Each of them is 
characterized by a choice of hidden and observable gauge group, a labeling of 
the fields and by their vev. In addition, one can analyze phenomenological 
properties and supersymmetric configurations ($F=D=0$) in this directory and 
determine the unbroken gauge group of a given vev-configuration.

For each orbifold model there are two standard vev-configurations: 
\code{StandardConfig1} and \code{Test\-Config1}. The first one cannot be 
changed, but the latter one can be and is used as default. In both 
configurations the full gauge group is selected as the observable sector and 
fields are labeled \code{F\_1}, \code{F\_2}, \code{F\_3}, $\ldots$, all with 
zero vev.

Note that the labels of the fields (see~\ref{app:fieldlabels}), sets of fields 
(see~\ref{app:sets}), monomials (see~\ref{app:monomials}), allowed couplings 
and mass matrices (created in the directory \code{/couplings>}) are saved in a 
vev-configuration. Hence, these data can only be accessed in the 
vev-configuration where they have been defined. In addition, note that all 
configuration-editing commands are not available in the vev-configuration 
\code{StandardConfig1}.

\myvspace
\paragraph{Command \code{analyze config}} 
Automatically check whether the current vev-configuration of the orbifold model 
allows for vacua with Standard Model, Pati-Salam or $\SU{5}$ gauge group,
\begin{equation}
\SU{3}_\text{C}\x\SU{2}_\text{L}\x\U{1}_\text{Y}\;,\;\; 
\SU{4}\x\SU{2}_\text{L}\x\SU{2}_\text{R}\quad\text{or}\quad\SU{5}\;,
\end{equation}
respectively, three generations of quarks and leptons and vector-like exotics. 
In the case the \orbifolder{} is not able to identify one of these 
possibilities for the current  orbifold model one obtains the output \code{No 
vev-configuration identified}. Otherwise, corresponding new vacua will be 
created and convenient labels will be assigned to all matter fields (e.g. 
\code{q\_1}, \code{q\_2} and \code{q\_3} for the three generations of quark 
doublets).

The command allows for two optional parameters: \code{print SU(5) simple roots} 
to print the simple roots of an intermediate $\SU{5}$ group that has been used 
in order to identify the hypercharge generator and \code{Xgenerations} with 
\code{X}$ = 0,1,2,3,\ldots$ to specify the (net) number of generations.

\myvspace
\paragraph{Command \code{compute F-terms}} Compute the $F$-terms using the 
superpotential that was created for the currently used vev-configuration (in 
the directory \code{/couplings>}. The optional parameter \code{max order(X)} 
allows to set an upper limit \code{X} on the order of superpotential couplings.

\myvspace
\paragraph{Command \code{create config(ConfigLabel)}} Create a new 
vev-configuration. Optionally, one can specify the origin of the new 
vev-configuration using the parameter \code{from(AnotherConfigLabel)}. If this 
parameter is not used the origin of the new vev-configuration is the standard 
vev-configuration \code{StandardConfig1}.

\myvspace
\paragraph{Command \code{delete config(ConfigLabel)}} Delete the 
vev-configuration \code{ConfigLabel}.

\myvspace
\paragraph{Command \code{find D-flat(fields)}} Identify gauge invariant 
monomials of \code{fields} as solutions to the $D=0$ supersymmetry condition, 
see~\ref{app:monomials}. This command allows for two parameters: i) \code{with 
FI} to allow for monomials with non-vanishing anomalous $\U{1}$ charge in order 
to cancel the FI term and ii) \code{save to set(SetLabel)} to save those fields 
in a set called \code{SetLabel} that are involved in the new monomials.

\myvspace
\paragraph{Command \code{find unbroken gauge group}} Depending on the vev 
assignment specified in the currently used vev-configuration, identify broken 
and unbroken (Abelian and non-Abelian) gauge group factors. In addition, the 
\U1 charges of all fields are re-computed in the new \U1 basis. There is one 
optional parameter: \code{print info} to display some details.

\myvspace
\paragraph{Command \code{print gauge group}} Print the choice of observable and 
hidden sector of the currently used vev-configuration, where the hidden sector 
gauge group factors are marked by brackets, e.g. \code{[SU(4)]}.

\myvspace
\paragraph{Command \code{print configs}} Print an overview of all vacua defined 
for this orbifold model. The currently used vev-configuration is highlighted by 
an arrow \code{->} in front, e.g. \code{-> "TestConfig1"}.

\myvspace
\paragraph{Command \code{rename config(OldConfigLabel) to(NewConfigLabel)}} 
Change the name of a vev-confi\-gura\-tion from \code{OldConfigLabel} to 
\code{NewConfigLabel}.

\myvspace
\paragraph{Command \code{select observable sector: parameters}} Assign a choice 
of observable and hidden gauge groups in the current vev-configuration. 
Admissible \code{parameters} are:
\begin{itemize}
\item \code{gauge group(i,j,...)}, where the indices \code{i,j} $= 1,2,\ldots$ 
      refer to the different non-Abelian gauge group factors sorted as 
      displayed by \code{print gauge group}. The indices provided are chosen as 
      part of the observable sector.
\item \code{full gauge group} All non-Abelian group factors are assigned as 
      observable sector.
\item \code{no gauge groups} No non-Abelian group factor is assigned as part of 
      the observable sector.
\item \code{U1s(i,j,...)}, where the indices \code{i,j} $= 1,2,\ldots$ refer to 
      the different \U1 gauge symmetries. The indices provided are chosen as 
      part of the observable sector.
\item \code{all U1s} All \U1s are assigned as part of the observable sector.
\item \code{no U1s} No \U1 is taken for the observable sector.
\end{itemize}
For example, assuming that the gauge group is \E6\x\SU3\x\E8, the instruction
\code{select observable sector: gauge group(1,2)} selects \E6\x\SU3 as the 
observable and \E8 as the hidden gauge groups.

\myvspace
\paragraph{Command \code{use config(ConfigLabel)}} Change the currently used 
vev-configuration to \code{ConfigLabel}.

\myvspace
\paragraph{Command \code{vev(fields) = ...}} Change the vevs of \code{fields} 
to new values. For example, \code{vev(*) = 0} turns off the vev of all fields, 
\code{vev(SetA) = rand} assigns random vevs to the fields of the set 
\code{SetA} and \code{vev(n\_1) = 0.1} sets the vev of \code{n\_1} to $0.1$.

%%%%%%%%%%%%%%%%%%% NEW SUBSUBSECTION %%%%%%%%%%%%%%%%%%%

\subsubsection{The directory \code{/vev-config/labels>}}
\label{app:directorylabels}

In this directory one can define, for each vev-configuration, appropriate 
labels for the fields. The main commands are \code{print labels} and 
\code{create labels}. In both cases, a summary table of massless fields is 
printed, sorted by those representations and $\U{1}$ charges that belong to the 
observable sector of the currently used vev-configuration. The observable 
sector can be changed using the command \code{select observable sector:...} in 
the directory \code{/vev-config>}, see~\ref{app:directoryconfig}.

\myvspace
\paragraph{Command \code{change label(A\_i) to(B\_j)}} Change the label of the 
field \code{A\_i} to \code{B\_j}.

\myvspace
\paragraph{Command \code{assign label(Label) to fixed point(k,l,n1,n2,n3,n4,n5,
n6)}} Assign \code{Label} to the fixed brane/point specified by \code{(k,l,n1,
n2,n3,n4,n5,n6)}.

\myvspace
\paragraph{Command \code{create labels}} First, a summary table of massless 
fields is displayed. Then the user is asked to specify a name for each line of 
the table.

\myvspace
\paragraph{Command \code{load labels(Filename)}} Load labels from the file 
named \code{Filename}. This command is disabled in the web interface.

\myvspace
\paragraph{Command \code{print labels}} Print a summary table of the currently 
used labels displaying the gauge representations with respect to the observable 
sector only.

\myvspace
\paragraph{Command \code{save labels(Filename)}} Save the labels to the file 
named \code{Filename}. This command is disabled in the web interface.

\myvspace
\paragraph{Command \code{use label(i)}} Change the currently used labels to the 
\code{i}-th labeling.

\addtocontents{toc}{\protect\setcounter{tocdepth}{0}}

%%%%%%%%%%%%%%%%%%%%%%%%%%%%%%%%%%%%%%%%%%%%%%%%%%%%%%%%%%%%%%%%%%%%%%%%%%
%  Bibliography
%%%%%%%%%%%%%%%%%%%%%%%%%%%%%%%%%%%%%%%%%%%%%%%%%%%%%%%%%%%%%%%%%%%%%%%%%%
\addcontentsline{toc}{section}{References}
\bibliography{Orbifold}
\bibliographystyle{styles/utphys}

\end{document}